\documentclass[a4paper]{jpsj2} \makeindex

\title{Magnetic Properties and Metastable States in Spin-Crossover \\
Transition of Co-Fe Prussian Blue Analogues}
\author{
Yusuk\'{e}~Konishi$^{1)}$\footnote{E-mail~address:~konishi@spin.phys.s.u-tokyo.ac.jp},
Hiroko~Tokoro$^{1) 2)}$,
Masamichi~Nishino$^{3) 4)}$ and 
Seiji~Miyashita$^{1) 4)}$
}

\inst{$^1$Department of Physics, The University of Tokyo,\\
Hongo 7-3-1, Bunkyo-ku, Tokyo 113-8656, Japan\\
$^2$JSPS, 5-3-1 Koji-machi, Chiyoda-ku, Tokyo 102-0083, Japan\\
$^3$Computational Materials Science Center, National Institute for Materials Science,\\
Tsukuba 305-0047, Japan\\
$^4$CREST, JST, 4-1-8 Honcho Kawaguchi, Saitama 332-0012, Japan\\
}

\abst{
The combination of spin transitions and magnetic ordering provides an interesting structure of phase transitions in Prussian blue analogues (PBAs). 
To understand the structure of stable and metastable states of Co-Fe PBA, it is necessary to clarify free energy as a function of magnetization and the fraction of the high-temperature component. 
Including the magnetic interaction between high-temperature states, we study the magnetic phase transition of Co-Fe PBA in addition to spin transitions.
Here, we take into account the degeneracy changes due to charge transfer between Co and Fe atoms accompanying the spin transition. 
In this study, the charge transfer between Co and Fe atoms is explicitly taken into account and also the ferrimagnetic structure of Co-Fe PBAs is expressed in the proper way.
First, we found systematic changes in the structures of stable and metastable states as functions of system parameters using mean field theory. 
In particular, the existence of a metastable magnetic-ordered high-temperature state is confirmed at temperatures lower than that of the hysteresis region of spin transitions. 
Second, we found that the magnetic interaction causes complex ordering processes of a spin transition and a magnetic phase transition.
The effect of a magnetic field on the phase structure is also investigated and we found metamagnetic magnetization processes.
Finally, the dynamical properties of this metastable state are studied by Monte Carlo method.
}
\kword{spin transition, charge transfer system, metastable structure, ferrimagnetism, metamagnetism}

\begin{document}
\maketitle
\section{Introduction}
In some octahedral coordinate iron complexes, a spin crossover (SC) may occur between a high-spin (HS) state and a low-spin (LS) state. An SC transition is induced by external stimulations such as temperature, photoirradiation, and magnetic field.\cite{1,2,3,4,5,6,7,8,9,10,11,12,13,14,15,16,17,18,19,20,21,22,23,24,25,26,27,28,29,30} 
The HS state is favorable at high temperature owing to its high degeneracy. 
On the other hand, the LS state is favorable at low temperature because it has a low energy with a low degeneracy.
Interesting transitions between HS and LS states are provided by the competition between entropy gain and energy gain.
It has been pointed out that cooperative interactions are important for SC transitions.
According to the system parameters, smooth transition or discontinuous first-order phase transition occurs in the SC transition.\cite{1,2,3,4,15,16,17,18,19,19.5,20,21,22}
This situation is well described by the Wajnflasz-Pick (WP) model,\cite{15} which gave the theoretical basis of the mechanism of the cooperative transitions using an Ising model with degenerate states. 
The WP model and its extended models have explained successfully the static and dynamical properties of various cases of SC transitions.\cite{19,19.5,20,21,22}
Control between the HS and LS states has been realized by photoirradiation as light-induced excited spin state trapping (LIESST),\cite{7,8,9,10,11,12,19,19.5,20,21,22,23} and the structure of the photoinduced metastable state of the systems has become an important topic.\cite{11,12,13,14}

Recently, we have reported that the metastable state can exist intrinsically at low temperature, which has been supported experimentally.\cite{13,14}
The effects of applying a magnetic field have been studied in some SC complexes.\cite{24,25,26,27,28,29,30}
Since the magnetic moment in the HS state is larger than that in the LS state, the HS state is stabilized by applying a magnetic field. 
A shift in transition temperature under an applied magnetic field was confirmed experimentally.\cite{24,25,26}
Using a pulsed high magnetic field, the creation of the HS state from the LS branch in the thermal hysteresis loop was observed and these results were interpreted using an Ising-like model.\cite{27,28,29,30}
However, since both the HS and LS states in the SC complexes are paramagnetic, the study of the effects of applying a magnetic field has been performed only for the paramagnetic region.

Some of the Prussian blue analogues (PBAs) \cite{31,32,33,34,35,36,37,38,39,40,41,42,43,44,45,46,47} show thermal-induced phase transitions and photomagnetic effects.\cite{38,39,40,41,42,43,44} 
In such materials, a strong cooperativity operates since transition-metal ions are bridged by CN ligands with a three-dimensional network structure.
Co-Fe PBA is an attractive material owing to its two-way photoswitching, i.e., magnetic $\leftrightarrow$ nonmagnetic.\cite{38,39,40,41}
The thermal induced phase transition between two states, i.e., Co(II)(HS, $S = 3/2$)-Fe(III)(LS, $S = 1/2$) and Co(III)(LS, $S =0$)-Fe(II)(LS, $S = 0$) is called a charge-transfer-induced spin transition (CTIST) \cite{41} in Co-Fe PBA.
Since the state of Co(II)-Fe(III) is favorable at high temperature owing to its high degeneracy, we call it a high-temperature (HT) state. 
Since the Co(III)-Fe(II) state is favorable at low temperature owing to its low energy, we call it a low-temperature (LT) state. 
In this material, the magnetic coupling between Co and Fe ions is antiferromagnetic. 
Therefore, the HT state is ferrimagnetic.

In this work, we study the effects of magnetic interaction and magnetic field on the cooperative phenomena of spin-crossover transitions and magnetic ordering for a Co-Fe PBA system using an extended Wajnflasz model. 
The calculations are performed using the Mean Field (MF) approximation and Monte Carlo (MC) methods. 
In particular, we study the effects for the metastable HT state ,which exists in low temperature region. 

The structure of this paper is as follows. In \S~2, the model used in this study is explained. 
In \S~3, the MF approximation used in this study is explained.
In \S~4, the effects of magnetic interaction and applying a magnetic field are presented using the MF approximation. 
The calculation results using the MC method are presented in \S~5. 
Section 6 is devoted to Summary and Discussion.

\begin{figure}[htbp]
  \begin{center}
    \includegraphics[height=40mm]{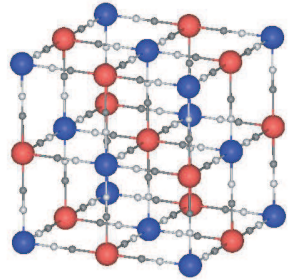}
  \end{center}
  \caption{Schematic illustration of present model of Co-Fe PBA.  Large blue circles are Co ions and red circles are Fe ions. Small white circles are N atoms and gray circles are C atoms. } 
  \label{CoFe}
\end{figure}

\section{Model}
Let us consider the CTIST phenomenon of Co-Fe PBA. This material consists of a bipartite lattice.
One of the sublattices is occupied by Fe ions and the other is occupied by Co ions (Fig.~1). 
We call a site of the Fe sublattice an A-site, and a site of the Co sublattice a B-site. 
The LT state consists mainly of Fe(II) and Co(III) ions.
The HT state consists mainly of Fe(III) and Co(II) ions. 
Here, we consider the degeneracy of the spin degree of freedom. 
At an A-site, Fe(II) in the LT state is $S = 0$, whose degeneracy is 1, and Fe(III) in the HT state is $S = 1/2$ with a degeneracy of 2. 
At a B-site, the degeneracy of $n_{\rm B}=0$ is 1 and that of $n_{\rm B}=1$ is 4 because Co(III) in the LT state is $S = 0$, and Co(II) in the HT state is $S = 3/2$ with a degeneracy of 4.
These degeneracies due to the spin states are listed in Table~\ref{table1}.
\begin{table}[htbp]
\begin{center}
\begin{tabular}{ccccc} 
state & site & spin state & $n$ & degeneracy\\
\hline
HT & A: Fe(III) & LS ($S=1/2$) & 0 & 2 \\
HT & B: Co(II)  & HS ($S=3/2$) & 1 & 4 \\    
LT & A: Fe(II)  & LS ($S=0$)   & 1 & 1 \\  
LT & B: Co(III) & LS ($S=0$)   & 0 & 1 \\
\hline
\end{tabular}
\end{center}
\caption{Spin state and degeneracy in Co-Fe PBA}
\label{table1}
\end{table}
Besides the spin degeneracy, the system has a degeneracy due to vibrational motions: $g_{\rm phonon}$(HS) and $g_{\rm phonon}$(LS).
The degeneracy due to phonons is much larger than that of spin.
However, in this study, we mainly use spin degeneracy, which is enough to provide the general aspects of the phase structure.

This charge transfer phenomenon is expressed by an electron transfer between Fe and Co atoms. 
To express this transfer, we introduce the quantity $n$, which is 1 for Fe in the LT state (i.e., $n_{\rm A}=1$) and 0 in the HT state ($n_{\rm A}=0$). 
Correspondingly, it is 0 for Co in the LT state ($n_{\rm B}=0$) and 1 in the HT state ($n_{\rm B}=1$). 
Here, $i$($j$) denotes A-(B-)sites. 
In this system, electrons are transferred between A and B sites.
We have introduced the following Hamiltonian for this system:
\begin{equation}
	{\cal H}=D\sum _{j}n_{\rm{B}}^{j}+\epsilon\sum_{\langle i,j \rangle}
n_{\rm{A}}^{i}n_{\rm{B}}^{j}
+4J\sum_{\langle i,j \rangle}s_{\rm{A}}^{i}s_{\rm{B}}^{j}
-2h\left(\sum_{i}s_{\rm A}^{i}+\sum_{j}s_{\rm B}^{j}\right).
\label{ham1}
\end{equation}
Here, $D$ ($>0$) is the difference in on-site energy between A- and B-sites, $\epsilon$ represents the repulsive interaction between the electrons at nearest-neighbor sites, and $J$ represents the magnetic interaction between A- and B-sites, and $s$ denotes the spin. 
The external magnetic field is given by $h$. 
Since an Fe and Co pair has an electron, the following relation holds:
\begin{equation}
     \sum_{i}n_{\rm{A}}^{i}+\sum_{j}n_{\rm{B}}^{j}=N,
     \label{napnb}
\end{equation}
where $N$ is half of the total number of Co and Fe ions of the lattice.
Here, we also introduce the magnetic states of the spins by $s_{\rm A}$ and $s_{\rm B}$.
When $n_{\rm A}=1$, $S=0$ and thus $s_{\rm A}$ takes only 0, and when $n_{\rm A}=0$, $S=1/2$ and thus $s_{\rm A}$ takes $-1/2$ or 1/2.
Similarly, when $n_{\rm B}=0$, $S=0$ and $s_{\rm B}$ takes 0, and when $n_{\rm B}=1$, $S=3/2$ and $s_{\rm B}$ takes $-3/2,-1/2,1/2$ or 1/2.
Specifying $s_{\rm A}$ and $s_{\rm B}$, the degeneracy is taken into account naturally.
In this model, electrons tend to stay at A-sites energetically, which denotes the LT state. 
When all the electrons are at A-sites, there is no energy cost due to $D$. 
We define this state as the perfect LT state. 
At high temperature, electrons tend to stay at B-sites owing to the difference in degeneracy mentioned above, which corresponds to the HT state.

Let us briefly review the properties of this type of Hamiltonian in the case with neither magnetic interactions nor a magnetic field ($J=h=0$).\cite{13}
In this case, it is known that the critical temperature ($T_{\rm 1/2}$) of charge order, at which $\langle n_{\rm B}\rangle=1/2$, is exactly given by
\begin{equation}
k_{\rm B}T_{\rm 1/2}=\frac{D}{\ln{\frac{g_{\rm HT}}{g_{\rm LT}}}},
\label{tc}
\end{equation}
where $k_{\rm B}$ is the Boltzmann constant.
$g_{\rm HT}$ and $g_{\rm LT}$ are the degeneracies of the HT (= 8) and LT (= 1) states, respectively.
With these $g_{\rm HT}$ and $g_{\rm LT}$, we have 
\begin{equation}
k_{\rm B}T_{\rm 1/2}=D/\ln{8}\simeq 0.4809D.
\label{model-tc}
\end{equation}

At this temperature, $\langle n_{\rm B}\rangle$ changes discontinuously in the first-order phase transition, when $T_{1/2}$ is larger than a critical value $T_{\rm Ising}/4$, where $T_{\rm Ising}$ is the critical temperature of the Ising model on the same lattice. 
On the other hand, if $T_{\rm Ising}/4<T_{1/2}$, $\langle n_{\rm B}\rangle$ changes smoothly.
On the simple cubic lattice, $k_{\rm B}T_{\rm Ising}=4.51\epsilon$.
Using this value, whether the transition is smooth or discontinuous depends on the ratio $\epsilon/D$, and the critical ratio is
\begin{equation}
\frac{\epsilon _{\rm c}}{D}=\frac{4}{4.51\ln8}\simeq 0.4266.
\label{ec1}
\end{equation}

In the mean field theory, $T_{\rm Ising MF}=6\epsilon$, and we have 
\begin{equation}
\frac{\epsilon _{\rm c}}{D}=\frac{4}{6\ln8}\simeq 0.3206,
\label{ec-mft}
\end{equation}
which is different from eq.~(\ref{ec1}).
Furthermore it has been pointed out that there are other critical values of $\epsilon$. 
When $\epsilon$ exceeds a critical value
\begin{equation}
\epsilon_{\rm CG}/D=1/6, 
\end{equation}
a low-temperature metastable branch appears. 
Moreover, when $\epsilon$ is larger than a critical value 
\begin{equation}
\frac{\epsilon_{\rm CX}}{D}=\frac{1}{6}
\frac{1+\sqrt{g_{\rm HT}/g_{\rm LT}}}{1-\sqrt{g_{\rm HT}/g_{\rm LT}}},
\end{equation}
a local stability of the HT state remains down to $T=0$, i.e., the low-temperature limit of the hysteresis loop reaches $T=0$.

\section{Mean Field Approximation}
In this section, we present a mean field formulation of the system with external magnetic interaction ($J$) and the applied magnetic field ($h$).
To study the model in the grand canonical ensemble, we extend the Hamiltonian including the chemical potential $\mu$.
\begin{equation}
	{\cal H}=D\sum _{j}n_{\rm{B}}^{j}+\epsilon
\sum_{\langle i,j \rangle}n_{\rm{A}}^{i}n_{\rm{B}}^{j}
+4J\sum_{\langle i,j \rangle}s_{\rm{A}}^{i}s_{\rm{B}}^{j}
-2h\left(\sum_{i}s_{\rm{A}}^{i}+\sum_{j}s_{\rm{B}}^{j}\right)
-\mu\left(\sum_{i}n_{\rm{A}}^{i}+\sum_{j}n_{\rm{B}}^{j}\right).
\label{hamj}
\end{equation}
In the mean field theory, the Hamiltonian is approximated as
\begin{eqnarray}
	{\cal H}_{\rm MFT}=(6n_{\rm B}\epsilon-\mu)\sum_i {n_{\rm A}^i}
+[6n_{\rm A}\epsilon+D-\mu]\sum_i {n_{\rm B}^i}+(24J s_{\rm A}-2h)\sum_i {s_{\rm B}^i} \nonumber\\
+(24J s_{\rm B}-2h)\sum_i {s_{\rm A}^i}
-6N\epsilon n_{\rm B}n_{\rm A}-24NJs_{\rm A} s_{\rm B},
\label{hammft}
\end{eqnarray} 
where $n_{\rm A}$, $n_{\rm B}$, $s_{\rm A}$ and $s_{\rm B}$ are the averages of quantities described as 
\begin{eqnarray}
n_{\rm A}=\frac{1}{N}\sum_{i}\langle n_{\rm A}^i\rangle,\quad n_{\rm B}=\frac{1}{N}\sum_{j}\langle n_{\rm B}^j\rangle \nonumber\\
s_{\rm A}=\frac{1}{N}\sum_{i}\langle s_{\rm A}^i\rangle,\quad s_{\rm B}=\frac{1}{N}\sum_{j}\langle s_{\rm B}^j\rangle.
\end{eqnarray}
The partition function is given as
\begin{eqnarray}
Z &=& {\rm Tr}\exp{(-\beta {\cal H}_{\rm MFT})}  \nonumber\\
&=& (e^{-\beta(6n\epsilon-\mu)}+2\cosh{(\beta(12Js_{\rm B}-h))})^N  \nonumber\\ 
&\quad& \times (1+2e^{-\beta[6n_{\rm A}\epsilon+D-\mu]}
[\cosh{(\beta(12Js_{\rm A}-h))}+\cosh{(3\beta(12Js_{\rm A}-h))}])^N.
\end{eqnarray} 
Here, it should be noted that Tr denotes the summation over all states, $(n_{\rm A}^i, s_{\rm A}^i)=(1,0),(0,-1/2),(0,1/2)$ and $(n_{\rm B}^j, s_{\rm B}^j)=(0,0),(1,-3/2),(1,-1/2),(1,1/2),(1,3/2)$.

Using the Hamiltonian (\ref{hammft}), we have
\begin{equation}
\langle n_{\rm B}^j\rangle=n_{\rm B}=\frac{2e^{-\beta(6n_{\rm A}\epsilon+D-\mu)}
[\cosh(\beta(12Js_{\rm A}-h))+\cosh(3\beta(12Js_{\rm A}-h))]}
{1+2e^{-\beta(6n_{\rm A}\epsilon+D-\mu)}[\cosh(\beta(12Js_{\rm A}-h))+\cosh(3\beta(12Js_{\rm A}-h))]},
\label{nb}
\end{equation}
and
\begin{equation}
\langle n_{\rm A}^i\rangle=n_{\rm A}=\frac{e^{-\beta(6n_{\rm B}\epsilon-\mu)}}
{e^{-\beta(6n_{\rm B}\epsilon-\mu)}+2\cosh(\beta(12Js_{\rm B}-h))}.
\label{na}
\end{equation}
Here, because of the condition for the electron number in eq.~(\ref{napnb}), $n_{\rm A}+n_{\rm B}=1$.
Hereafter, we express the HT fraction 
\begin{equation}
n=n_{\rm B},
\end{equation}
and thus 
\begin{equation}
n_{\rm A}=1-n.
\end{equation}
Then, we have 
\begin{eqnarray}
\frac{2e^{-\beta[6(1-n)\epsilon+D-\mu]}[\cosh(\beta(12Js_{\rm A}-h))+\cosh(3\beta(12Js_{\rm A}-h))]}
{1+2e^{-\beta[6(1-n)\epsilon+D-\mu]}[\cosh(\beta(12Js_{\rm A}-h))+\cosh(3\beta(12Js_{\rm A}-h))]} \nonumber\\
+\frac{e^{-\beta(6n\epsilon-\mu)}}{e^{-\beta(6n\epsilon-\mu)}
+2\cosh(\beta(12Js_{\rm B}-h))}
=1,
\end{eqnarray}
from which we determine the chemical potential $\mu$.
Setting $f=e^{\beta\mu}$, we obtain
\begin{equation}
f=\sqrt{\frac{\cosh(\beta(12Js_{\rm B}-h))}{\cosh(\beta(12Js_{\rm A}-h))+\cosh(3\beta(12Js_{\rm A}-h))}}
e^{\beta(\frac{D}{2}+3\epsilon)}.
\label{fug}
\end{equation}

The free energy $F$ is given by
\begin{eqnarray}
F &=& -\frac{T}{N}\ln Z \nonumber\\
&=&-T\ln{(fe^{-6\beta\epsilon n}+2\cosh(\beta (12Js_{\rm B}-h)))} \nonumber\\
&\quad&
-T\ln{[1+2fe^{-\beta(6\epsilon(1-n)+D)}(\cosh(\beta (12Js_{\rm A}-h))+\cosh(3\beta (12Js_{\rm A}-h)))]} \nonumber\\
&\quad&-6\epsilon n(1-n)-24Js_{\rm A}s_{\rm B}.
\end{eqnarray}

The self-consistent equations for $s_1$ and $s_2$ are obtained by the conditions of $\partial F/\partial s_{\rm A} =\partial F/\partial s_{\rm B} =0$.  
The condition $\partial F/\partial s_{\rm A} =0$ leads to 
\begin{equation}
	s_{\rm B}=-\frac{\sinh(\beta(12Js_{\rm A}-h))+3\sinh(3\beta(12Js_{\rm A}-h))}{\cosh(\beta (12Js_{\rm A}-h))+\cosh(3\beta (12Js_{\rm A}-h))}n,
	\label{sb}
\end{equation}
and from $\partial F/\partial s_{\rm A} =0$, we have
\begin{equation}
	s_{\rm A} =-\tanh(\beta (12Js_{\rm B}-h))n.
	\label{sa}
\end{equation}
The magnetization of the system is given by 
\begin{equation}
m=s_{\rm A}+s_{\rm B}.
\end{equation}
To obtain the thermal properties of the system, we have to find solutions that satisfy the eq.~(\ref{nb}),(\ref{sb}) and (\ref{sa}).

\section{Phase Transition in Mean Field Theory}
\begin{figure}[htbp]
$$
  \begin{array}{cc}
	  \mbox{\includegraphics[height=80mm,angle=-90]{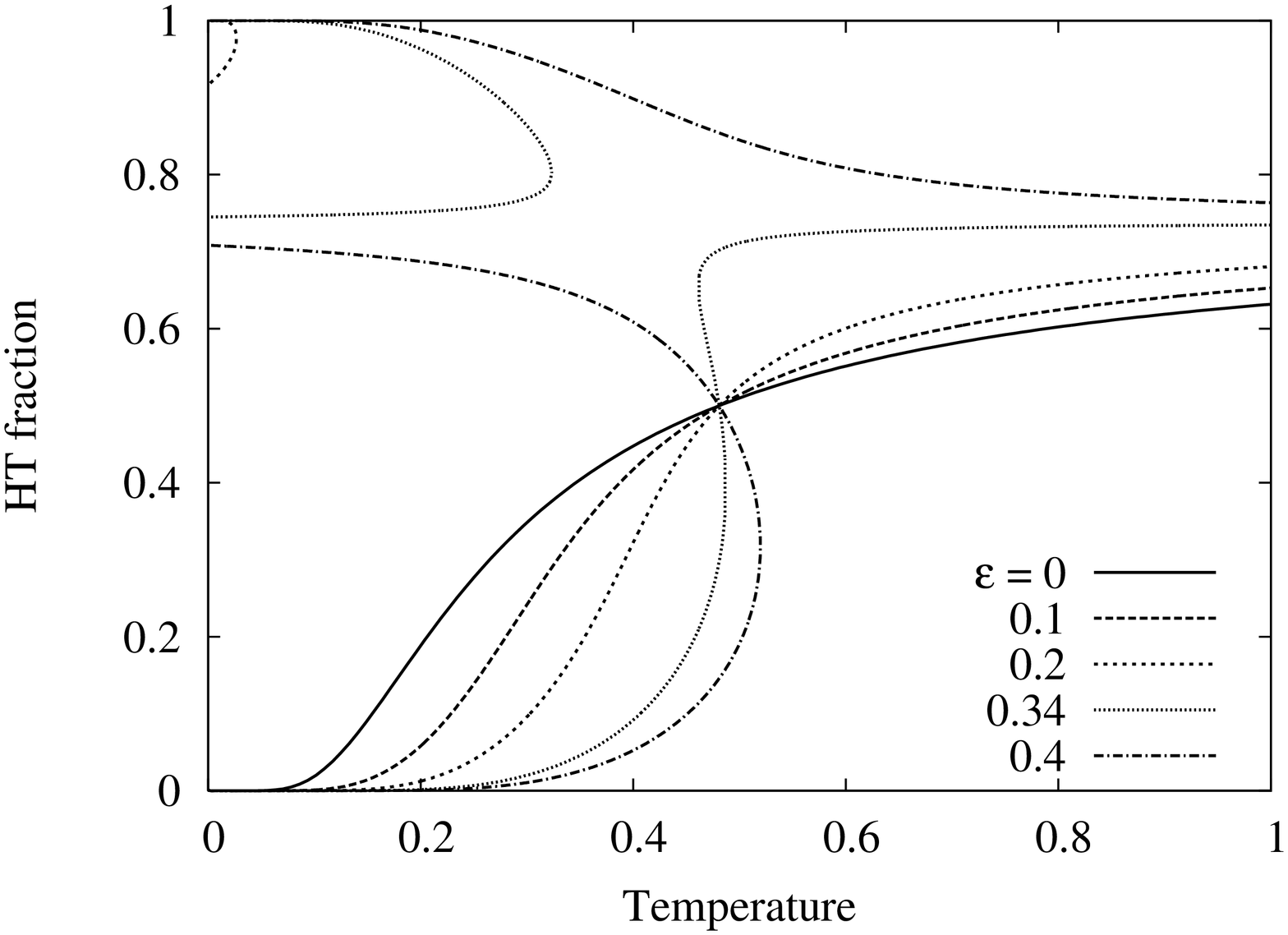}}
    &
    \mbox{\includegraphics[height=80mm,angle=-90]{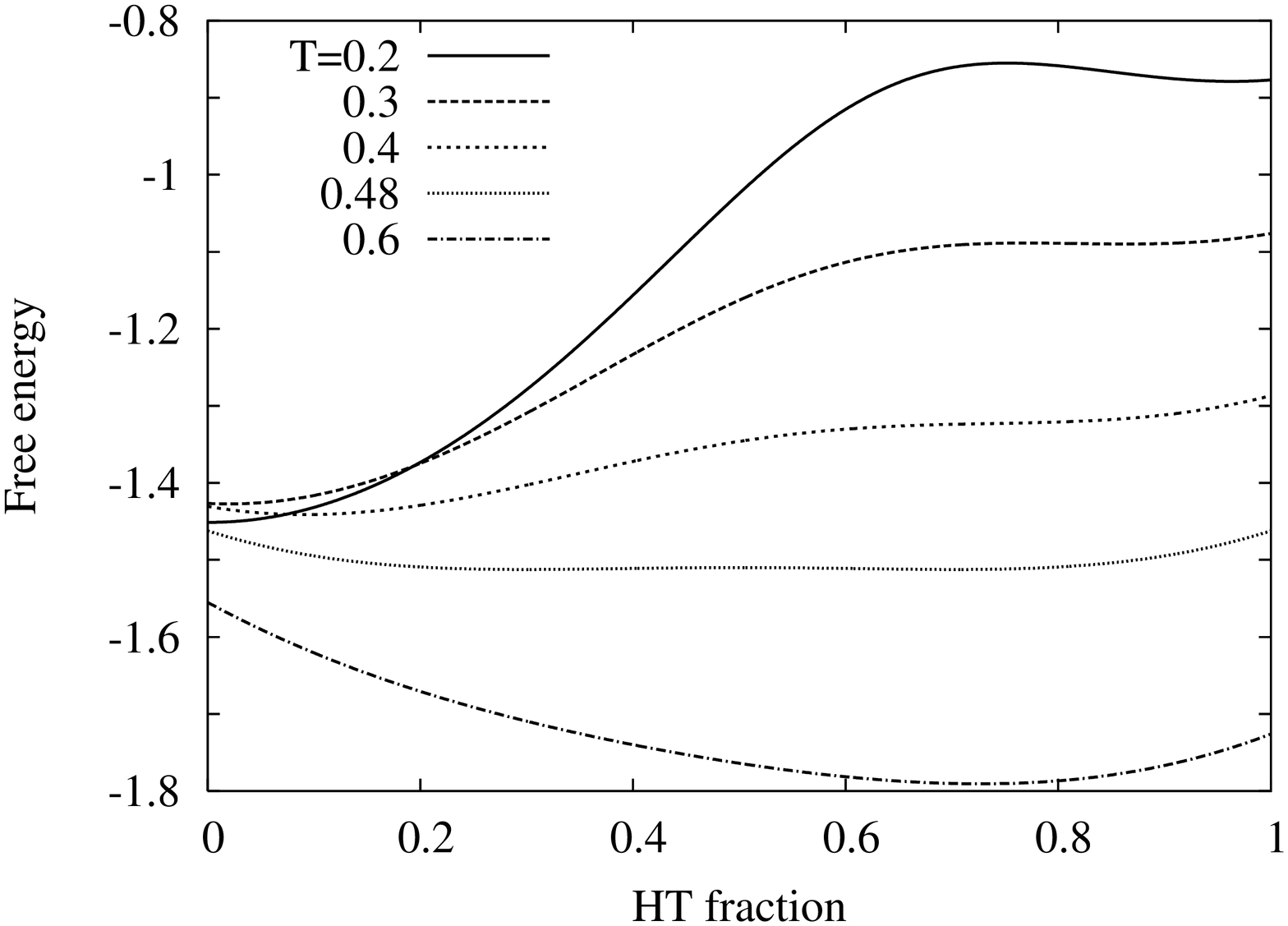}}\\
    {\rm (a)}&{\rm(b)}\\
  \end{array}
$$
  \caption{(a)~Temperature dependence of HT fraction with parameters of $J=0,D=1$ and $\epsilon=0,0.1,0.2,0.34$ and 0.4.
  (b)~Free energy curves as functions of HT fraction.} 
  \label{fig-mft-j0}
\end{figure}

Hereafter, we take $D$ as the unit of energy.
First, we review the properties of the system without magnetic interaction $(D,J,h)=(1,0,0)$ as a reference.
Let us study the effect of repulsive interaction, $\epsilon$.
The temperature dependences of the HT fraction ($n(T)$) obtained by solving eq.~(\ref{nb}) with eq.~(\ref{fug}) are presented in Fig.~\ref{fig-mft-j0}(a) for the values of $\epsilon = 0, 0.1, 0.2, 0.34$, and 0.4. 
When $\epsilon = 0$, the transition between HT and LT states is gradual and continuous. 
Until the critical value of $\epsilon(=0.32D)$, eq.~(\ref{ec-mft}), the transition is continuous. 
When $\epsilon > \epsilon_{\rm c}$, the transition becomes discontinuous and it shows a first-order phase transition.
In this case, the transition temperature ($T_{\rm c}$) is $T_{1/2}=0.48$ (eq.~(\ref{model-tc})), and the system has three solutions at $T_{\rm c}$.
The solutions with the largest and smallest HT fractions are stable states, whereas that with the intermediate HT fraction represents an unstable state.
For $\epsilon_{\rm CG}=1/6<\epsilon<\epsilon_{\rm CX}\simeq0.349$, the HT state is the metastable state, as shown in Fig.~\ref{fig-mft-j0}(a).
For large values of $\epsilon$ ($\epsilon>\epsilon_{\rm CX}$), the local minimum of the HT state exists at all temperatures as shown in Fig.~\ref{fig-mft-j0}(a) ($\epsilon=0.4$).
In Fig.~\ref{fig-mft-j0}(b) the corresponding free energy profiles for $\epsilon=0.34$ are depicted for various temperatures, where we find the metastable structure at low temperature.

\subsection{Effect of magnetic interaction}
\begin{figure}[htbp]
  $$
  \begin{array}{cc}
      \mbox{\includegraphics[keepaspectratio=true,width=80mm]{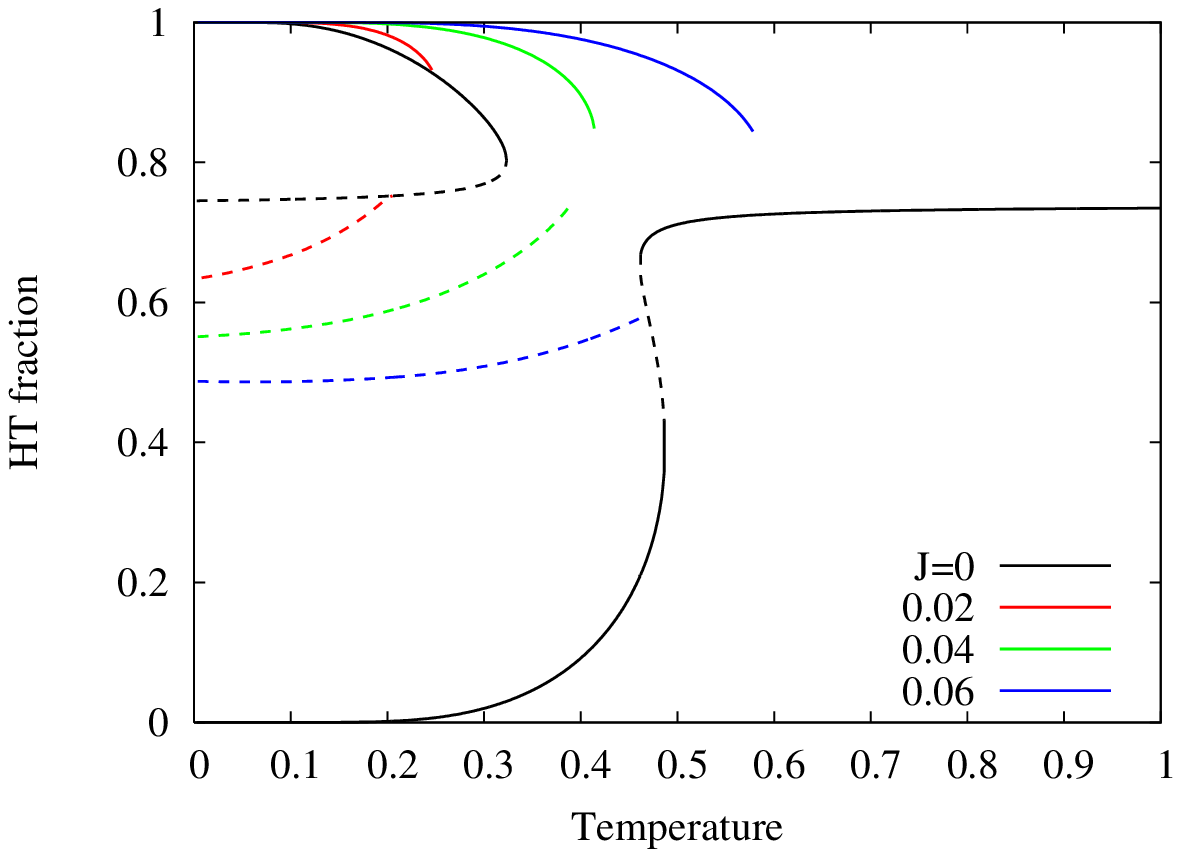}}&\mbox{\includegraphics[keepaspectratio=true,width=80mm]{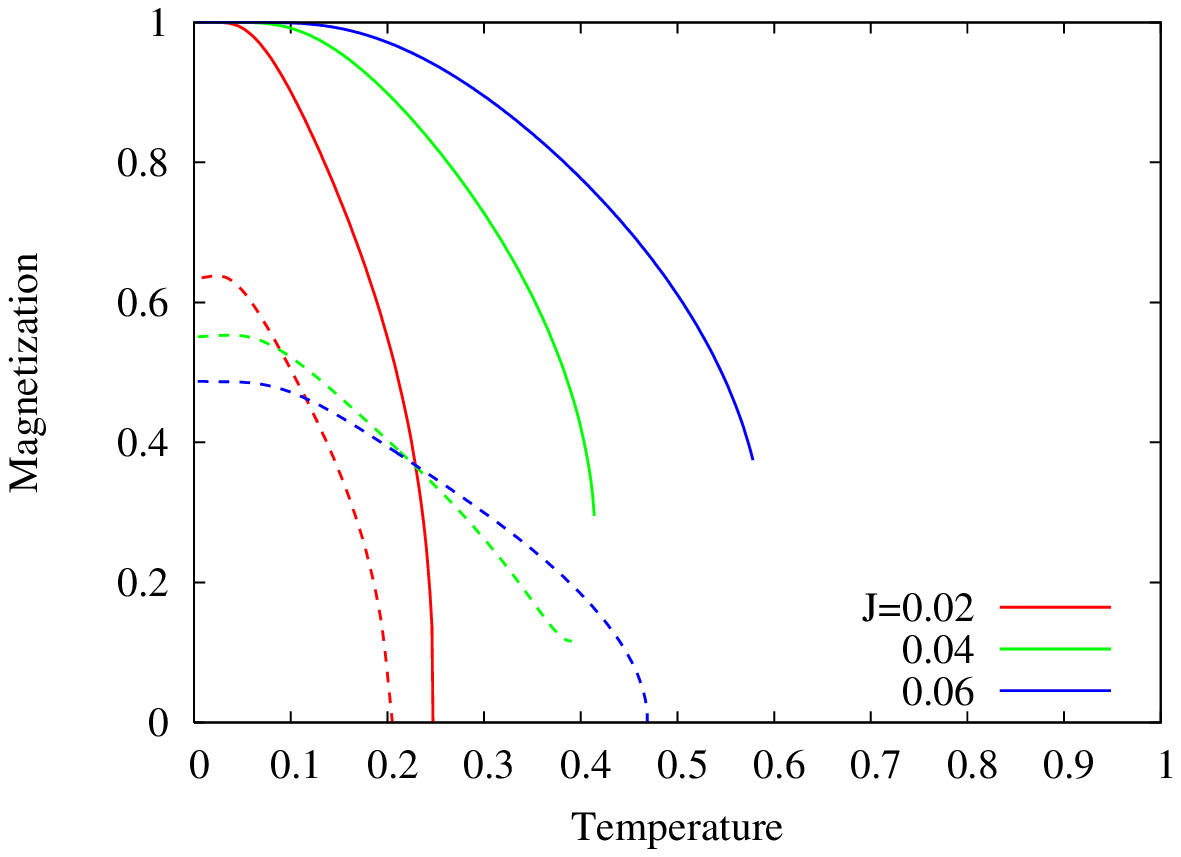}}\\
      {\rm (a)}&{\rm (b)}\\
  \end{array}
  $$
  \caption{(a)~Temperature dependence of HT fraction for 
  $\epsilon=0.34$, $D=1$ and $J=0,0.02,0.04$ and 0.06. 
  (b)~Temperature dependences of magnetization 
   for $\epsilon=0.34, D=1,$ 
  and $J=0.02, 0.04$ and 0.06.
  The solid curves represent stable and metastable solutions and the 
  dashed curves represent unstable solutions.}
  \label{fig-mft-a}
\end{figure}

Let us consider the effect of magnetic interaction using $J$ without the external magnetic field ($h = 0$). 
The temperature dependences of $n$ are depicted in Fig.~\ref{fig-mft-a}(a) with the parameters ($D,\epsilon , h$) = (1, 0.34, 0) with $J$ = 0, 0.02, 0.04, and 0.06, where the system possesses a HT metastable branch at low temperature in the case without the magnetic interaction ($J=0$). 
When the magnetic interaction is included, we find that the metastable branch is enlarged.
In the case of $J = 0.02$, the HT fraction of the metastable HT state is increased. 
The HT fraction of the unstable solution is also affected.
Here, it should be noted that the solution of $J=0$ is not changed unless the magnetization has nonzero solution in the MF treatment.
Indeed, the thermal hysteresis loop does not change at all in this case. 
When $J$ is increased to 0.04, the metastable branch is further enlarged and terminates at $T=0.41$, at which $n$ changes discontinuously to the solution of $J=0$ (black line), where the magnetization vanishes discontinuously.
We call this temperature $T_{\rm cmag}$.
The temperature dependences of magnetization $m(T)$ are depicted in Fig.~\ref{fig-mft-a}(b).
For $J=0.06$, the metastable branch terminates at $T=0.57$.
The magnetic states in the case $J=0.02$ and $J=0.04$ are metastable, whereas it is stable, i.e., the equilibrium state in the case $J=0.06$.
This means that the model is a simple ferrimagnetic without CTIST.

\begin{figure}[htbp]
$$
  \begin{array}{cc}
      \mbox{\includegraphics[keepaspectratio=true,height=60mm]{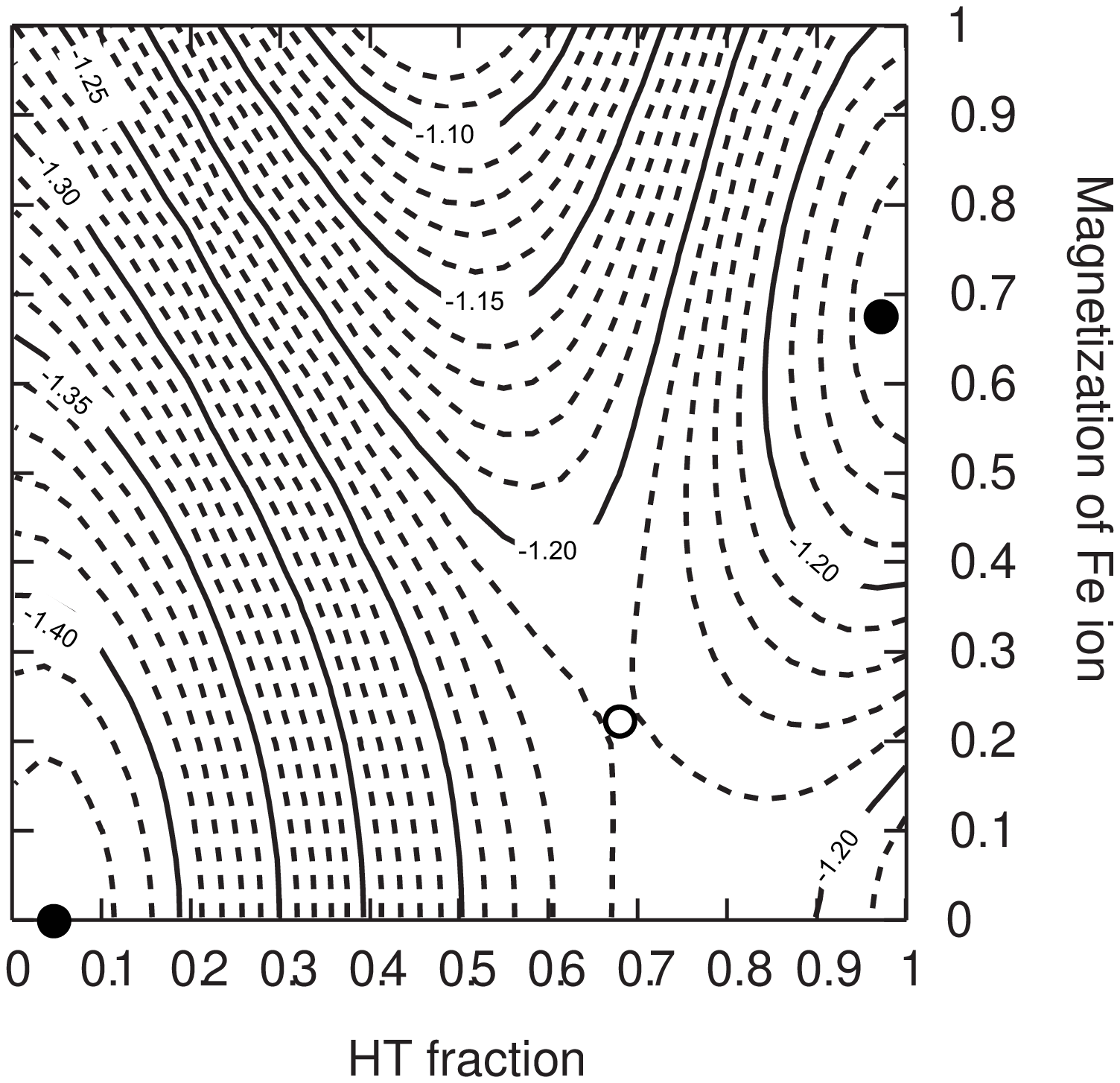}}&\mbox{\includegraphics[keepaspectratio=true,height=60mm]{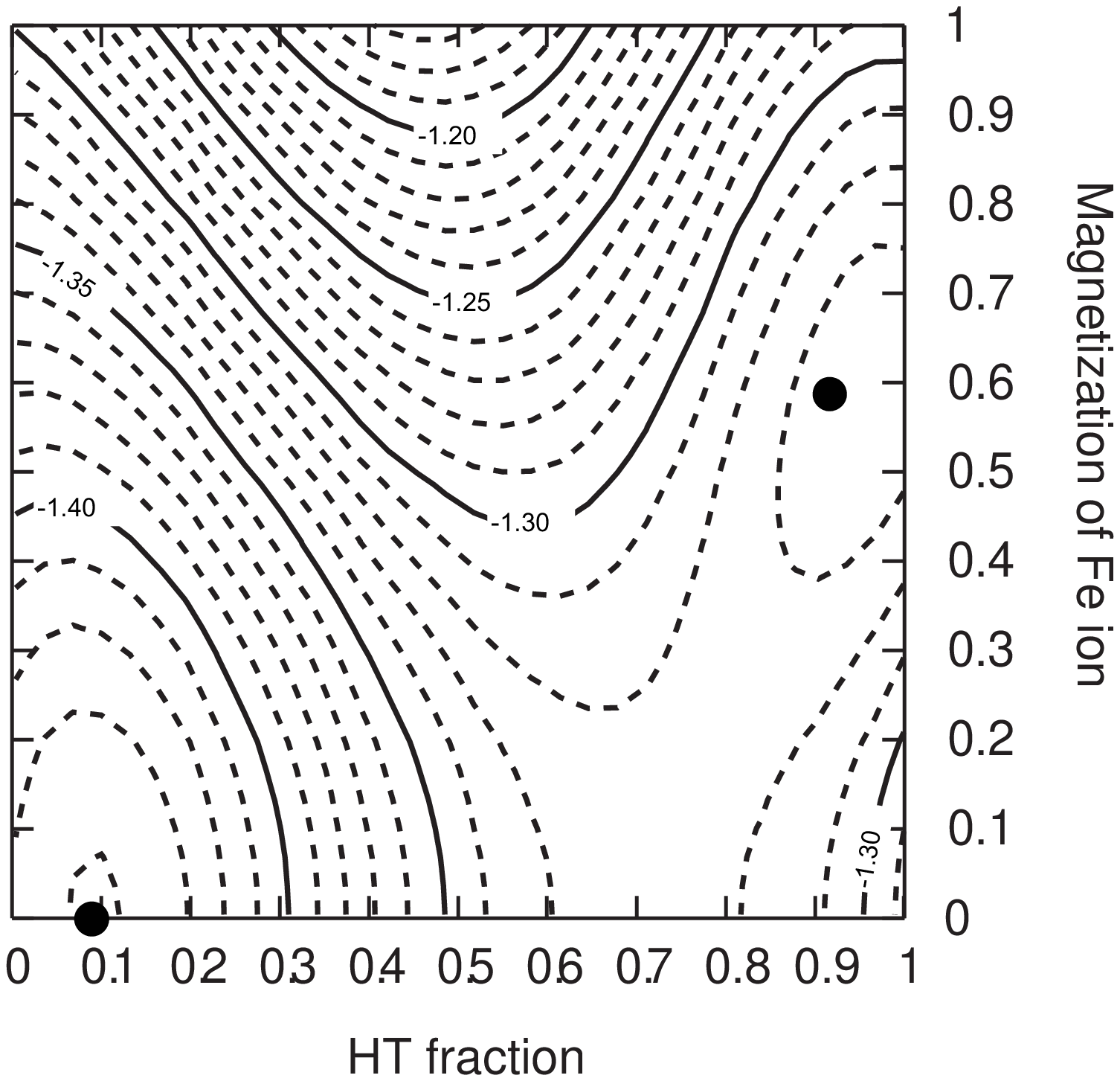}}\\
      {\rm (a)~T=0.35}&{\rm (b)~T=0.4}\\
      \\
      \mbox{\includegraphics[keepaspectratio=true,height=60mm]{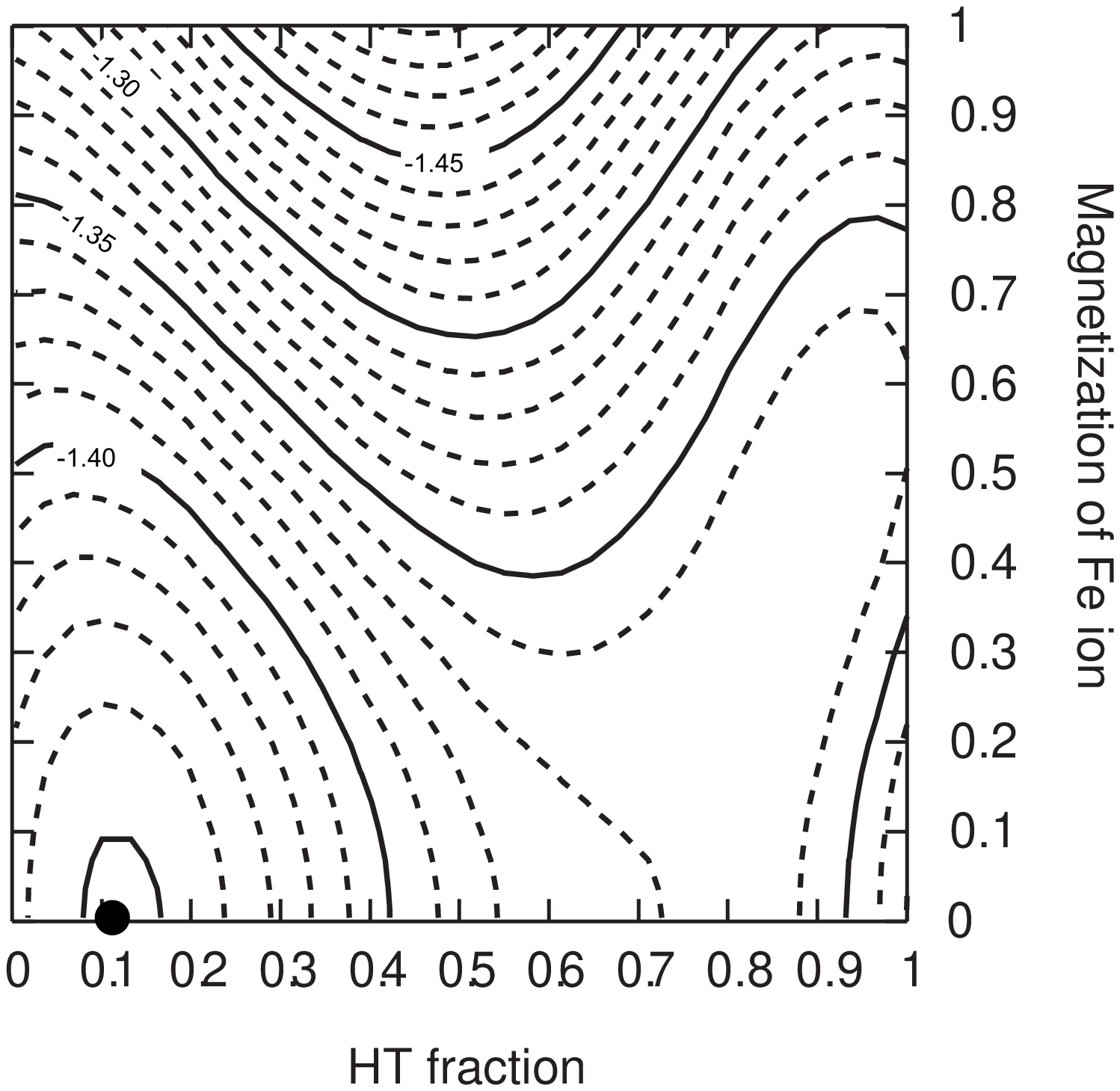}}\\
      {\rm(c)~T=0.42}\\
  \end{array}
  $$
  \caption{Contour plots of free energy at 
  $\epsilon=0.34$, $D=1$ and $J=0.04$.
  Closed circles indicate the local minimum points and the open circle indicates
  the saddle point.}
  \label{fig-fe}
\end{figure}

In Fig.~\ref{fig-mft-a}(a), we find that the metastable solution (solid green line) and the unstable solution (dotted green line) terminate at different temperatures $T=0.41$ and $T=0.39$, respectively, and are not connected to each other.
This is due to the fact that the present model has three order parameters, i.e., ($n,s_{\rm A},s_{\rm B}$).
At $T=0.35$, there are three solutions that are shown in Fig.~\ref{fig-fe}(a) by closed circles for local minimum points and an open circle for a saddle point.
When $T$ is increased up to $T=0.4$, the saddle point disappears, as shown in Fig.~\ref{fig-fe}(b), where two stable solutions correspond to points on the green solid line and black solid line denoting the LT state.
At $T=0.42$, the metastable point disappear and only one stable point remains, which gives a point on the black line. 
For large values of $J$, HT remains locally stable until $T=0$, as we see in Fig.~\ref{fig-mft-a}(a) for $J=0.06$.

\begin{figure}[htbp]
  \begin{center}
      \includegraphics[keepaspectratio=true,width=80mm]{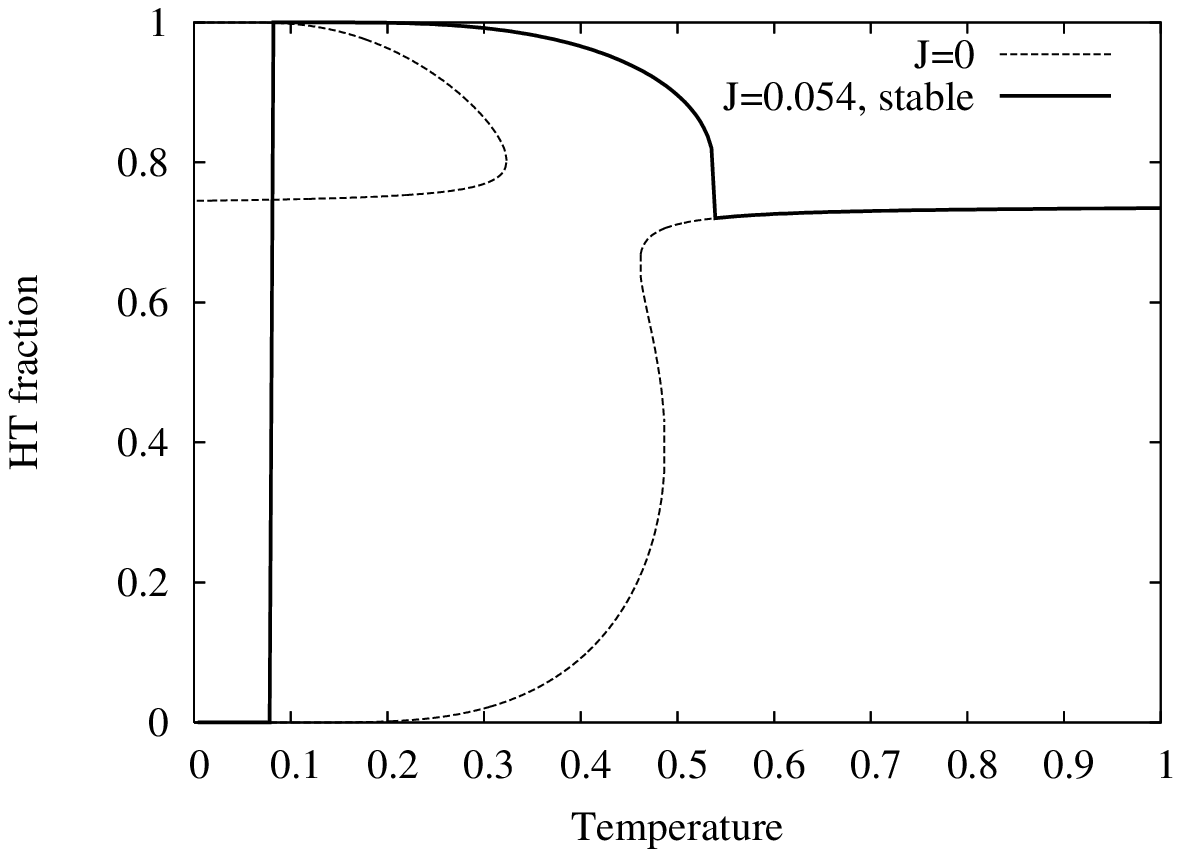}
  \end{center}
  \caption{Temperature dependence of HT fraction for $D=1$,
  $\epsilon=0.34$ and $J=0.054$.
  The dotted curves indicate solutions of the self-consistent 
  equations for $J=0$.
  The solid curve indicates the stable solution for $J=0.054$.}
  \label{fig-mft-b}
\end{figure}

Here, we point out a unique dependence of the stability of the ferrimagnetic state on $J$.
If $J<J_{\rm c}=D/18$, the ferrimagnetic state is metastable and not an equilibrium state at low temperatures.
Here, $J_{\rm c}$ is obtained by the comparison of the energy of the complete ferrimagnetic state and the LT state.
In Fig.~\ref{fig-mft-b}, we show an example of $J=0.54<J_{\rm c}$.
There, the equilibrium solution is drawn by a bold solid curve.

\subsection{Effect of external field}
\begin{figure}[htbp]
$$
\begin{array}{cc}
\mbox{\includegraphics[width=80mm]{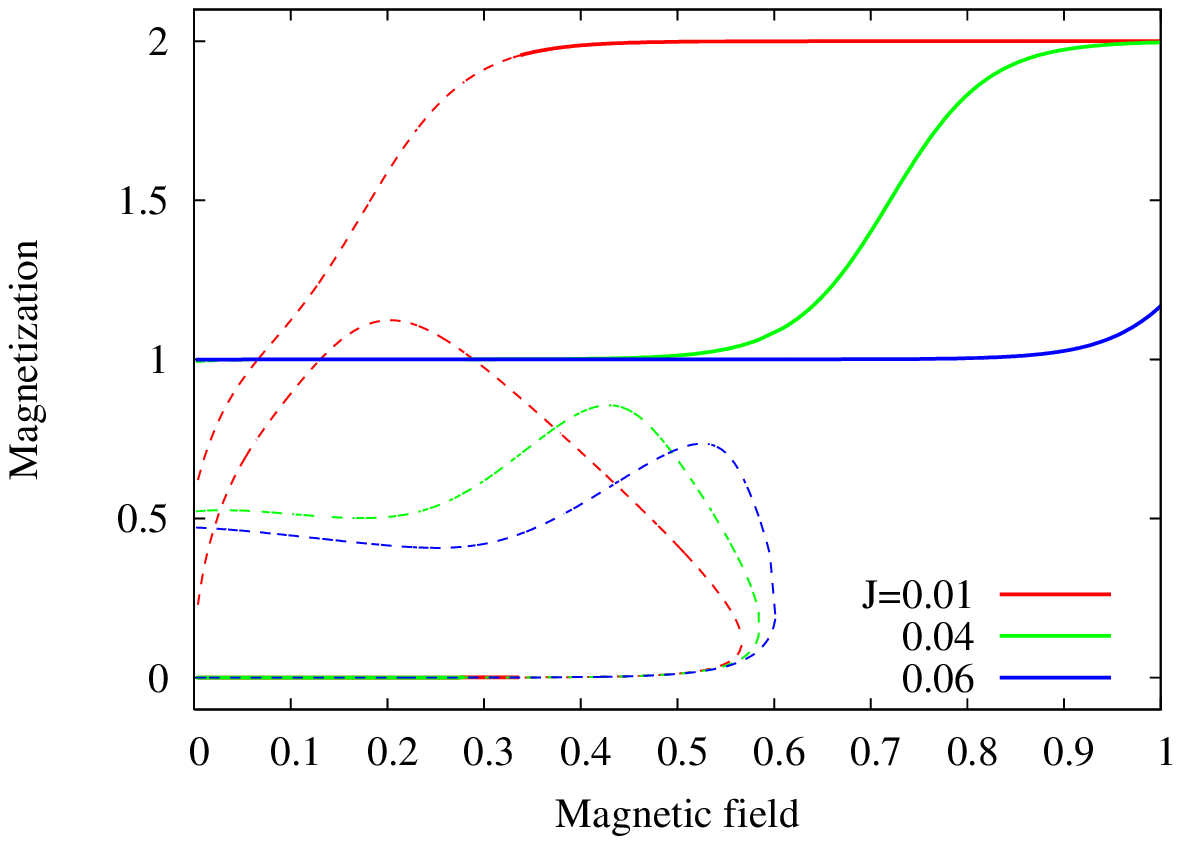}}
&
\mbox{\includegraphics[width=80mm]{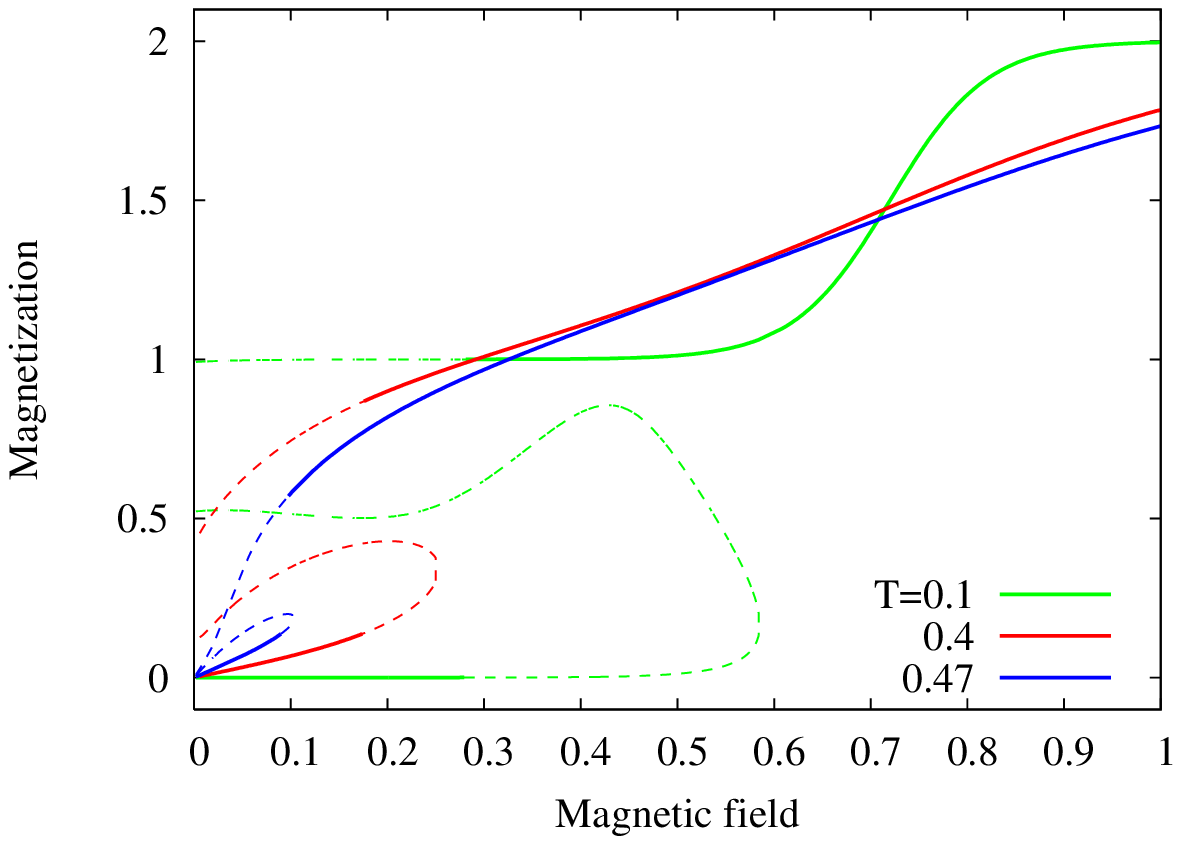}}\\
{\rm (a)}&{\rm(b)}
\end{array}
$$
\caption{Magnetization curves for (a)~$T=0.1$, $J=0.01,
0.04,0.06$, (b)~$J=0.04$, $T=0.1,0.4,0.47$.
The bold curves indicate the stable solutions and the dashed curves indicate 
the metastable and unstable solutions.}
\label{fig-h1}
\end{figure}

Let us consider the effect of the applied magnetic field $h$. 
Here, we study the case of $\epsilon=0.34$ where both thermal hysteresis and the HT metastable exist at low temperature.
The magnetic field dependence of magnetization at $T = 0.1$ is presented in Fig.~\ref{fig-h1}(a) for various values of $J$ ($J = 0.01, 0.04$, and 0.06), where the $T_{\rm cmag}$ values of $J = 0.01, 0.04$, and 0.06 are 0.133, 0.39, and 0.57, respectively. 
In the case of $J = 0.01$, the LT state is a stable state in the absence of a magnetic field. 
For a weak field, the system is always in the LT phase.
The magnetization is suddenly induced up to nearly the saturated value at $h=0.338$. 
The magnetization value of 2 indicates the ferromagnetic state of Fe ($S = 1/2$) and Co ($S = 3/2$) ions in the HT state. 
In the case of $J = 0.04$, a stable state in the absence of a magnetic field is still the LT state. 
At $h=0.281$, the magnetization is induced up to 1 and shows a plateau until $h\simeq0.45$.
After that, the magnetization is induced up to 2. 
The magnetization of 1 indicates an antiferromagnetic coupling between Fe and Co ions in the HT state, indicating that the plateau at the magnetization of 1 is the region of the ferrimagnetic state due to the antiferromagnetic interaction $J$. 
For the case of $J = 0.06$, the ground state is ferrimagnetic even at $h=0$.
The magnetization is induced up to 2 in a strong magnetic field.
These results indicate that the applied magnetic field can induce transitions between the three states: the LT state, the antiferromagnetic coupling HT state, and the ferromagnetic coupling HT state.

Next, we study the temperature dependence of magnetization.
Figure~\ref{fig-h1}(b) shows the magnetization with $J = 0.04$ at $T = 0.1, 0.4$, and 0.47. 
In the case of $T = 0.4$, the system is in the LT state in a small field and the magnetization gradually increases until $h=0.175$. 
In this field, the magnetization jumps to 0.87, and then gradually increases up to 2. 
In this case, the metastable state exists, indicating that the hysteresis loop can be observed. 
In the case of $T = 0.47$, the system is in the LT state in a small field and the magnetization increases up to 2 with a jump at $h=0.090$. 

\begin{figure}[htbp]
\includegraphics[width=80mm]{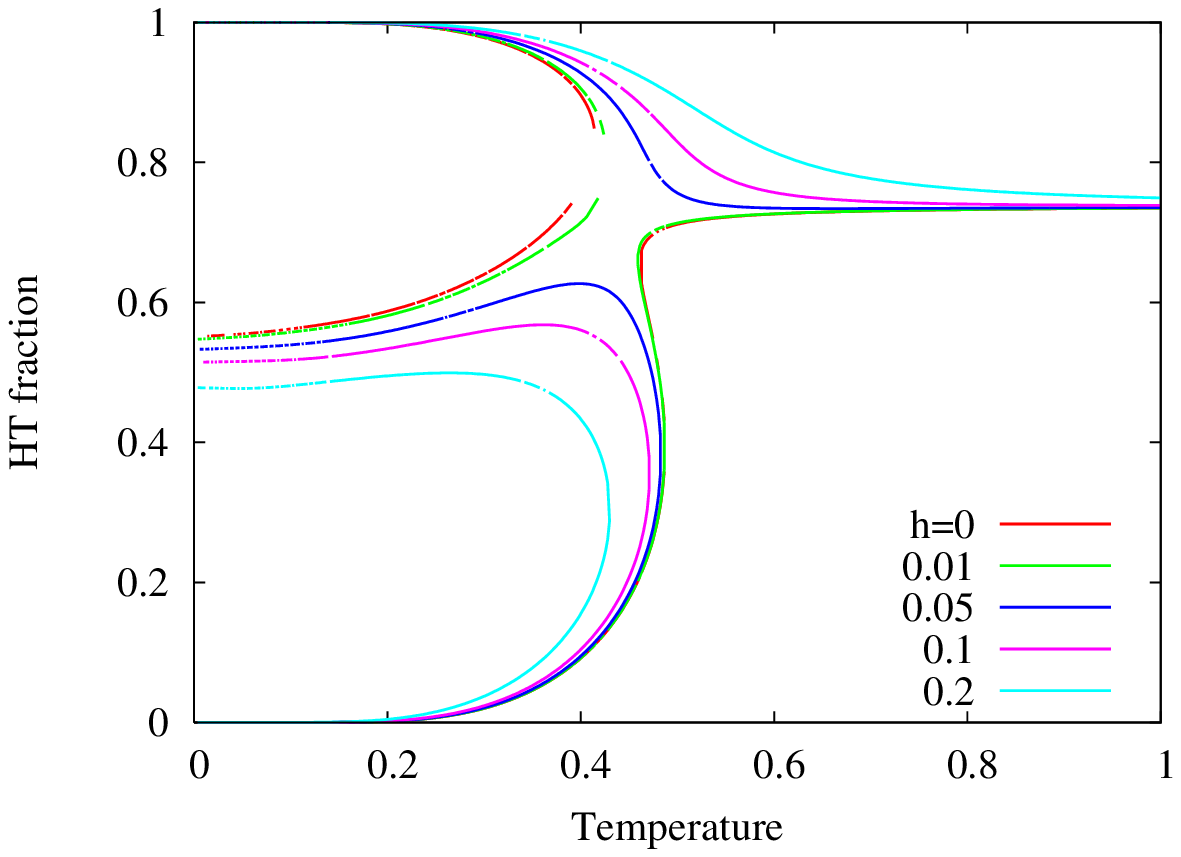}
\caption{Temperature dependences of HT fraction for $D=1$, $\epsilon=0.34$, $J=0.04$ and $h=0, 0.01, 0.05, 0.1$ and 0.2.}
\label{e034j004th}
\end{figure}

Now, we study the effect of the magnetic field on the structure of $n(T)$.
In Fig.~\ref{e034j004th}, $n(T)$ are presented for ($D, \epsilon, J$) = (1, 0.34, 0.04) and $h = 0, 0.01, 0.05, 0.1$, and 0.2. 
As $h$ increases, the HT fraction at high temperature increases. 
In the metastable HT state, the HT fraction is increased. 
This tendency is enhanced as $h$ increases. 
In contrast to that in the case of Fig.~\ref{fig-mft-a}(a) where $J$ changes, the HT fraction changes even above $T_{\rm cmag}$, in the present figure.

\section{Monte Carlo Simulations}
In the previous section, we study the static properties of the model using the mean field approximation, where the effect of short-range fluctuation is not taken into account.
Therefore, in this section, we study the model by Monte Carlo simulation and confirm the results obtained in the previous section.
Furthermore, we study dynamical properties and discuss the metastability of the model.

\subsection{Equilibrium states}
Here, we study the changes in $n(T)$ and $m(T)$ starting from the initial states of $n=1$ and $m=1$, respectively, which correspond to the photoinduced HT saturated state at very low temperature.
At each step, we performed 10000 MCSs for transient steps and 20000 MCSs to measure the physical quantities.
The system size is $16\times16\times16$, which is large enough to study thermal properties.
We heat the temperature up to $T=1$ and then cool it down to the original temperature.
We investigate $n(T)$ in the case of $(D,\epsilon)=(1,0.4)$ and $J=0,0.025,0.05$.
Results of these simulations are depicted in Fig.~8.
In the case of $J=0$, we found a smooth $n(T)$ in Monte Carlo simulation, whereas, in MF theory, we find that the HT phase exists down to $T=0$ as a metastable state that causes the first-order phase transition at some temperature (Fig.~\ref{fig-mft-j0}(a) $\epsilon=0.4$).
This fact indicates that in MC, different types of $n(T)$ appear from the MF result for the same set of parameters, which is naturally expected.
Nevertheless, in MC, we find a similar trend of $n(T)$ to that of MF (Fig.~\ref{fig-mft-a}(a)).
Here, it should be noted that in MC, we find $n(T)$ at high temperature, where the state is paramagnetic, is also caused by $J$. 
In the MF result, $n(T)$ did not change up to $m=0$. 
Indeed, when $J=0.025$, $n(T)$ changes from that for $J=0$, although the system is paramagnetic in both cases.
If we increase $J$ further, the system shows a first order phase transition.
This trend is the same as that in MF theory.

Because we sweep the temperature, the result depends on sweeping rate.
However, we find that the position of the transition from metastable HT to stable LT changes little with sweeping rate (Fig.~\ref{fig-mc2}), which suggests that the state is stable or metastable and is rather well defined by the system parameters.
\begin{figure}[htbp]
$$
  \begin{array}{cc}
    \includegraphics[keepaspectratio=true,width=80mm]{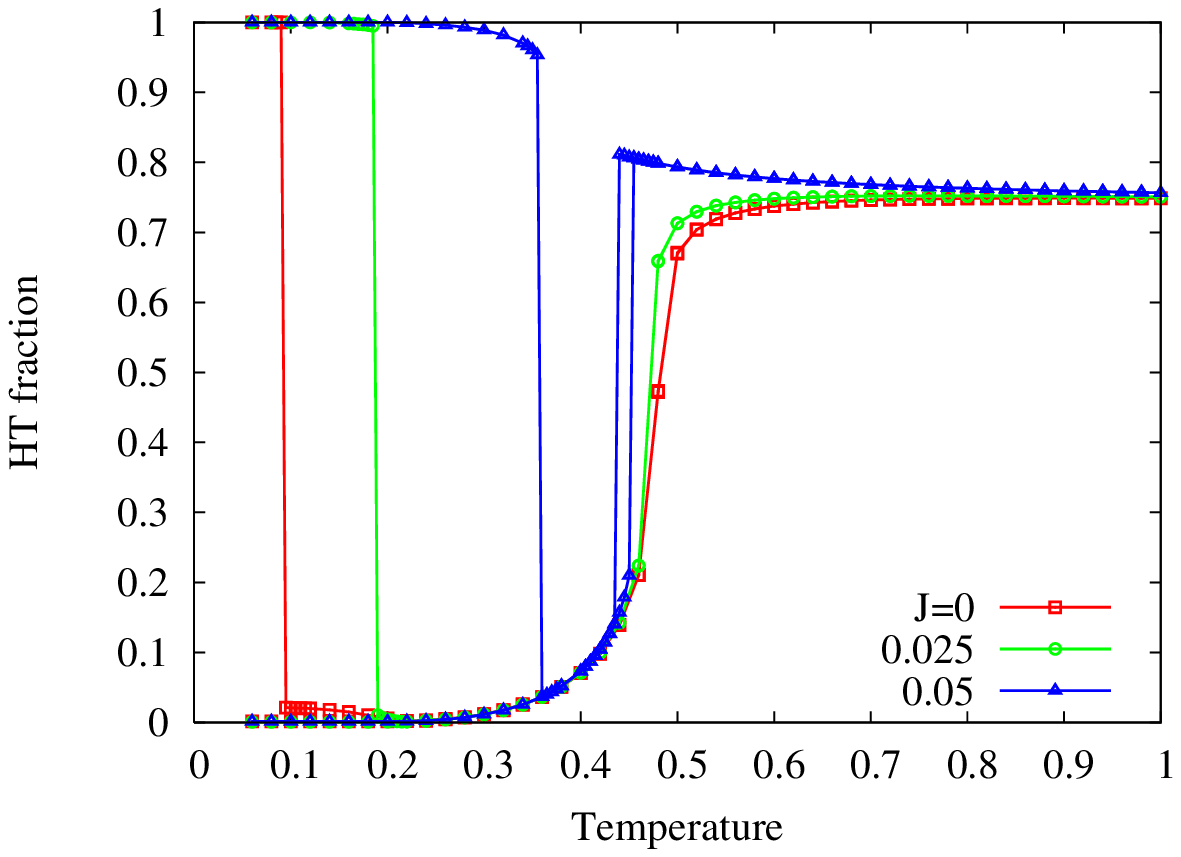}
      &
    \includegraphics[keepaspectratio=true,width=80mm]{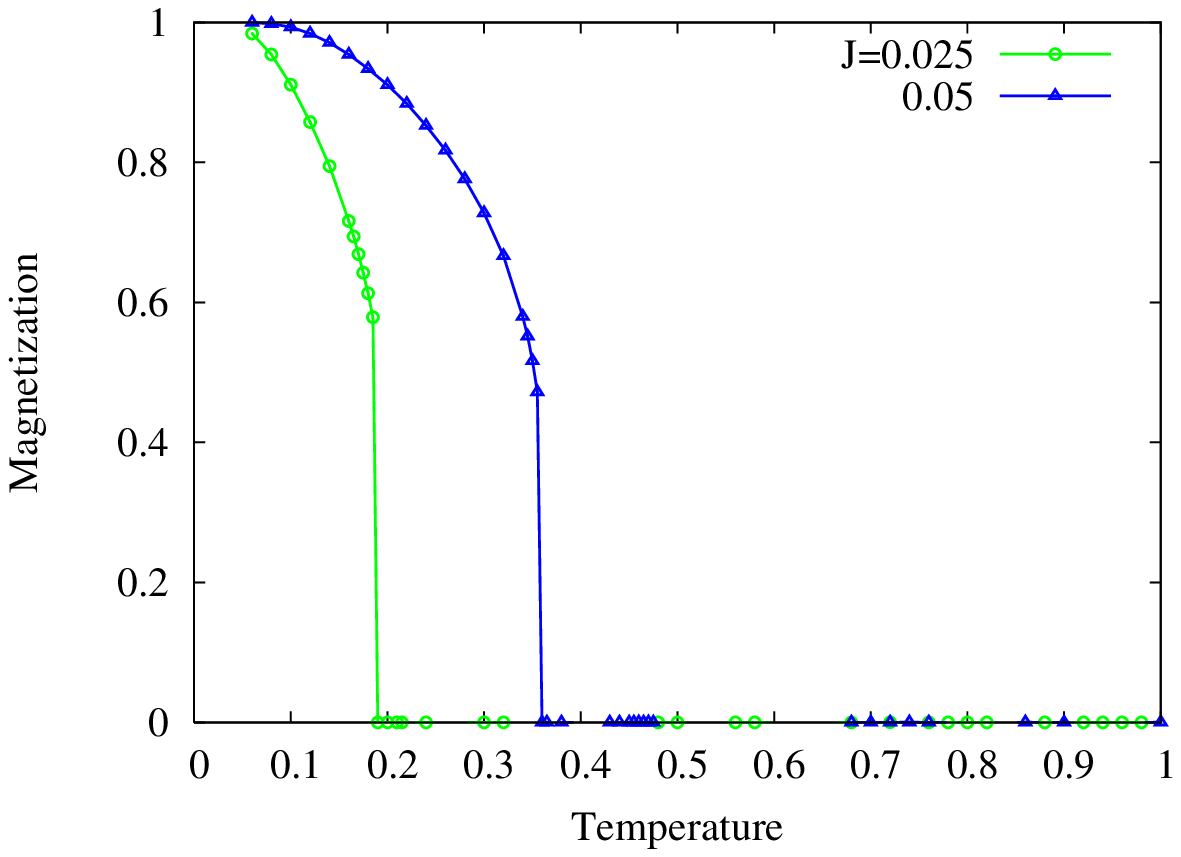}\\
      {\rm (a)}&{\rm (b)}\\
  \end{array}
$$
  \caption{MC simulations of temperature dependences of HT fraction and magnetization
  for $D=1$, $\epsilon=0.4$, and $J=0,0.025,0.05$.
  }
  \label{fig-mc1}
\end{figure}

\begin{figure}[htbp]
    \includegraphics[keepaspectratio=true,width=80mm]{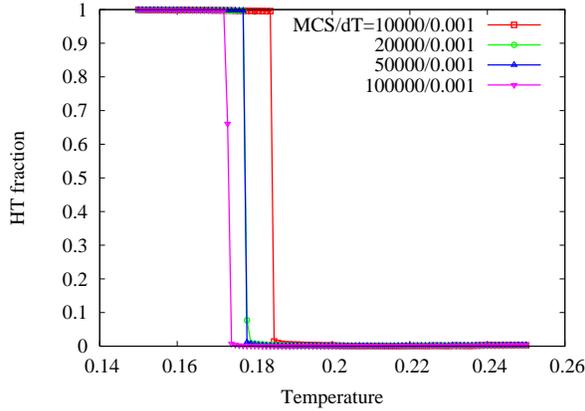}
  \caption{
  MC simulations of $n(T)$ for $D=1, \epsilon=0.4$ and $J=0.025$.
  We change sweeping rate from $10^4$ MCSs to $10^5$ MCSs.
  The transition temperature is around $T=0.18$ for all sweeping rates.
  }
  \label{fig-mc2}
\end{figure}

\subsection{Relaxation and evaluation of unstable states}
\begin{figure}[htbp]
  $$
  \begin{array}{cc}
    \mbox{\includegraphics[keepaspectratio=true,width=80mm]{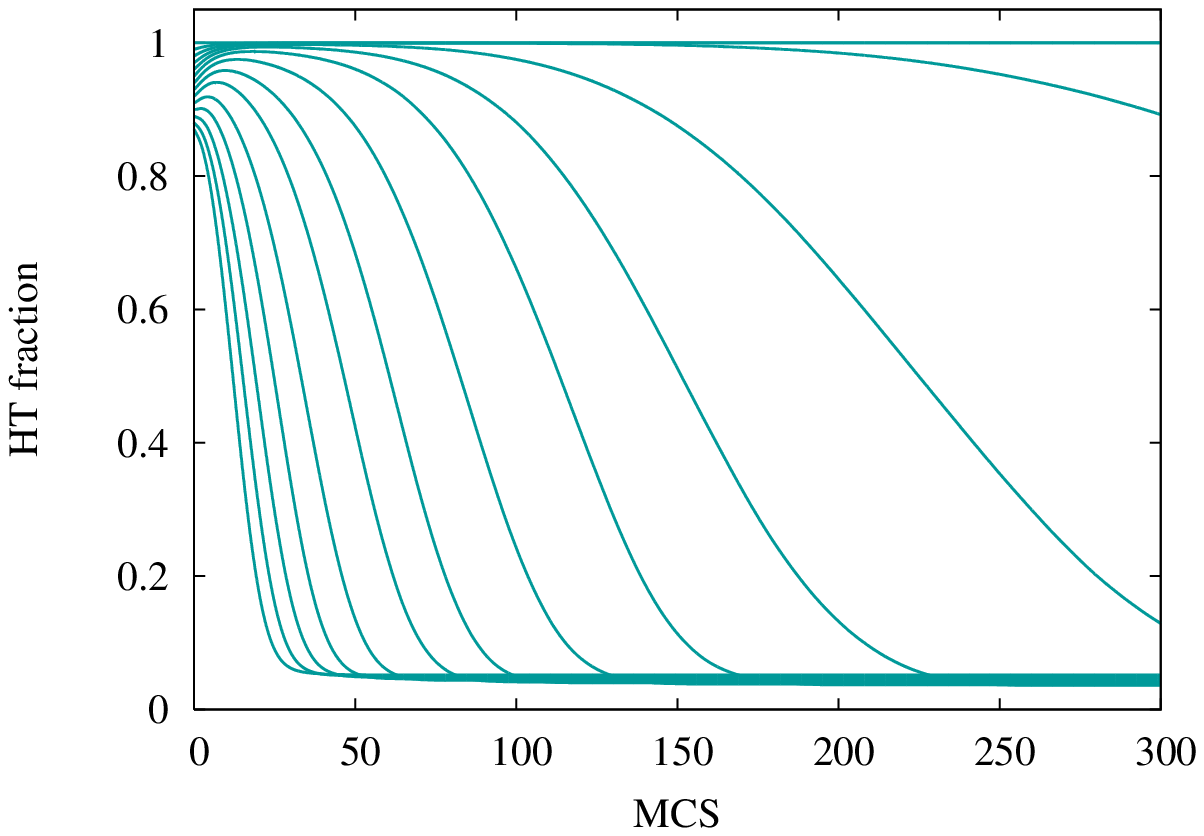}}
    &
    \mbox{\includegraphics[keepaspectratio=true,width=80mm]{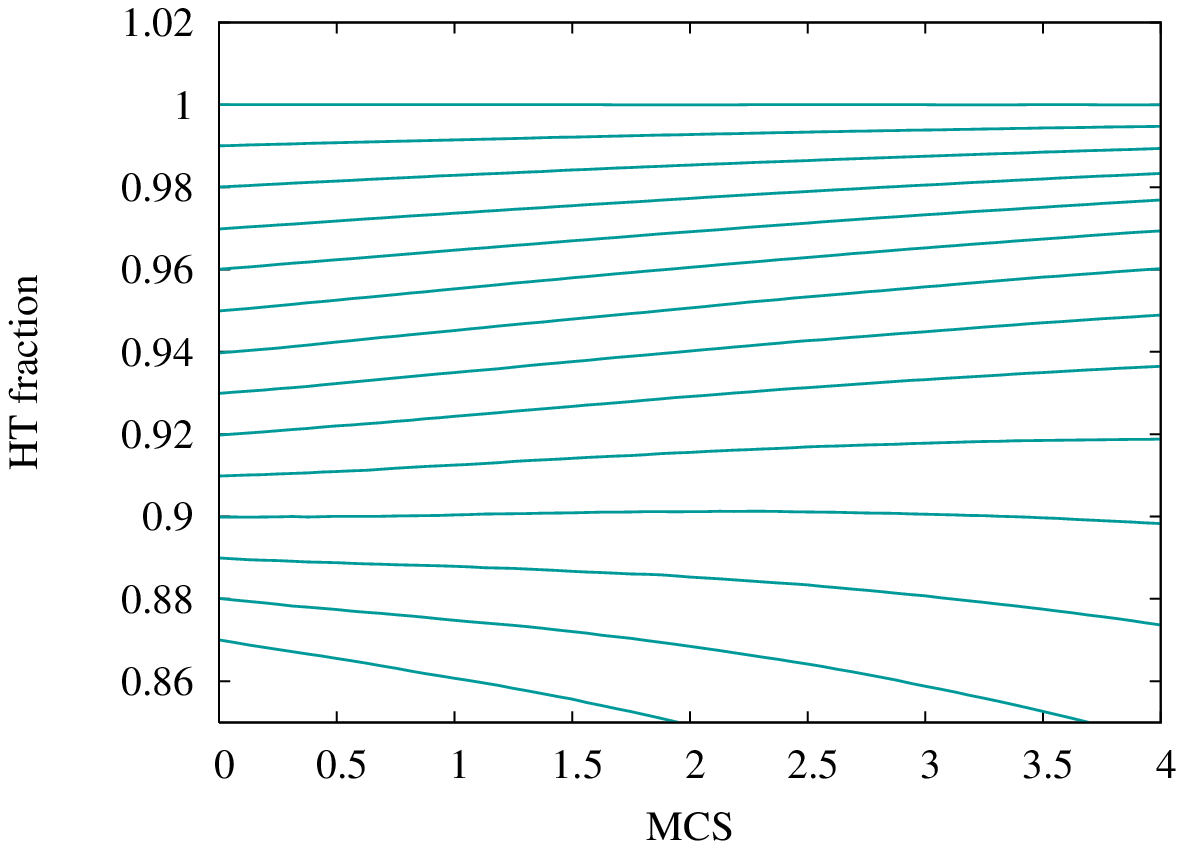}}\\
    {\rm (a)}&{\rm (b)}\\
  \end{array}
  $$
  \caption{Relaxation of HT fraction for $D=1$, $\epsilon=0.4$, $J=0.025$ and $T=0.1$.
  (a)~These simulations are started from $n=0.87,0.88,...,1.00$.
  (b)~Magnified figure for short time.
  }
  \label{fig-rlx}
\end{figure}

\begin{figure}[htbp]
    \includegraphics[keepaspectratio=true,width=80mm]{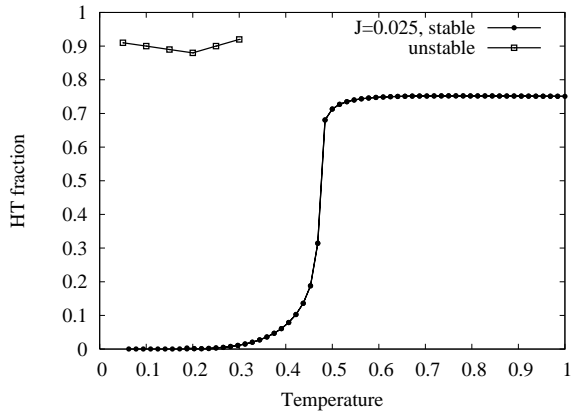}
  \caption{
  Stable and unstable points for $D=1,\epsilon=0.4$ and $J=0.025$.
  The squares indicate the unstable points determined by MC simulation.
  The circles indicate equilibrium points.
  }
  \label{fig-rlx-frac}
\end{figure}

Finally, we come to the structure of the metastable state, \cite{13} e.g., the local minimum of this metastable HT state exists at low temperature. 
The existence or nonexistence of the metastable phase is a significant issue in the field of photoinduced phase transition because whether the photoinduced HT phase is a metastable phase or a photogenerated thermodynamically unstable phase is of current interest. 
From this point of view, we study the relaxation processes from the metastable HT to LT states by MC simulation. 
Here, we adopt the initial state of $n\simeq1$ where the positions of HT sites are chosen randomly. 
Such a state can be produced by rapid cooling or light irradiation where the HT sites locate uncorrelatedly. 
Figure~\ref{fig-rlx}(a) shows the relaxation processes.
In this section, we study the system using ($D, \epsilon, J, h$) = (1, 0.4, 0.025, 0) at $T = 0.1$ for different HT fractions, e.g., 0.87, 0.88, ..., 1.00.
The system size is $128\times128\times128$.
When the HT fraction is smaller than 0.90, the relaxation curves show monotonic decreases. 
In contrast, when $n$ is larger than 0.90, the relaxation curves increase initially, as shown in Fig.~\ref{fig-rlx}(b). 
This is a particular phenomenon due to the existence of a local minimum of the HT state. 
After that, the states relax to the LT state with sigmoidal shapes.
Here, we find a separatrix at $n_{\rm sp}=0.89$ that denotes the position of an unstable point in MC. 
We study the position of $n_{\rm sp}$ at other temperatures and estimate the temperature dependences of unstable points, which are plotted in Fig.~\ref{fig-rlx-frac} as dots with a stable $n(T)$ dependence.
The stable and unstable points show a qualitatively similar temperature dependence to that obtained by MF.

\section{Summary and Discussion}
The spin states and magnetic properties of Co-Fe PBA were studied. 
We introduced a model in which we can take into account the fact that charge transfer occurs between Co and Fe ions that are ordered antiferromagntically at low temperatures. 
The effects of magnetic interaction and an external magnetic field on the structure of the metastable states of $n(T)$ and $m(T)$ were investigated by MF and MC  methods. 
We found a systematic change in the temperature dependence of $n(T)$ on the parameters of the system, which was found in our previous study. 
Namely, in that study, by changing the parameter $\epsilon$, we reproduced the same trend of $n(T)$. 
In this work, in particular, the combination of the ordering processes of $n(T)$ and $m(T)$ is studied. 
The magnetic interaction induces a magnetic order and enhances the ordering of $n(T)$. 
The low-temperature metastable HT branch is enlarged and even connected to the HT branch. 
Furthermore, we found a unique temperature dependence of the set of $(n(T),m(T))$. 
That is, the solution of $(n(T),m(T))$ disappears discontinuously, which causes a discontinuous transition of $n(T)$ and $m(T)$. 
The external magnetic field also causes a change in $n(T)$ in a similar sequence to that found for magnetic interaction. 
In our formulation, the ferrimagnetic structure is taken into account, and thus we found that the magnetization shows two steps, i.e., among the LT state, the ferrimagnetic state and the field-induced ferromagnetic state. 
The qualitative features of the effect of magnetic interaction obtained using the MF approximation were confirmed by MC simulation, although in the MC method we found that the magnetic interaction has an effect even in the paramagnetic states.
In order to characterize the low-temperature metastable branch, we studied the relaxation from the HT state. 
We found that the direction of initial relaxation can be upwards or downwards when the metastability exists depending on the initial $n$.
That is, $n$ relaxes monotonically to LT when $n$ is smaller than the threshold value $n_{\rm sp}$, whereas it first increases to its metastable value when $n > n_{\rm sp}$.
Studying the temperature dependence of $n_{\rm sp}$, we reproduced a temperature dependence similar to that of the unstable solution in the MF study.

The magnetic properties of PBA obtained in this study, particularly the trend of the changes in the temperature dependences of $n$ and $m$ on system parameters will be useful for classifying various materials, not only those belonging to Fe-Co PBA but also more general materials in which charge transfer and spin-crossover induce phase transitions.
In particular, in the present formalism, we can take into account details of microscopic structure changes, such as the spin values in the HT structure, and thus the present formalism will be useful for studying the combined phenomena of charge transfer, spin-crossover, and magnetic ordering in a wide range of materials.

In the present study, we use a short-range interaction $\epsilon$ as an elastic interaction approximately.
However, to be realistic, this elastic interaction is long-range.
In this type of materials, this long-range interaction is important to study the effect of volume change or that of pressure.
The effect of long-range elastic interaction will be reported elsewhere in the near future.

\section*{Acknowledgements}
This work is partially supported by Grant-in-Aid from the Ministry of Education, Culture, Sports, Science and Technology, and also by NAREGI Nanoscience Project, Ministry of Education, Culture, Sports, Science and Technology, Japan. 
The authors also thank the Supercomputer Center, Institute for Solid State Physics, the University of Tokyo for the use of facilities.

%\bibliographystyle{unsrt}
%\bibliography{pba}

%\input{bib.tex}

\end{document}